\documentclass[12pt]{iopart}
\usepackage{times}
\usepackage{mathptmx}
\usepackage{graphicx}

\newcommand{\kbf}{{\mathbf{k}}}
\newcommand{\dd}{{\mathrm{d}}}
\newcommand{\ee}{{\mathrm{e}}}
\newcommand{\Id}{{\mathrm{Id}}}
\newcommand{\calC}{{\mathcal{C}}}
\newcommand{\counit}{{\varepsilon}}

\renewcommand{\i}[1]{{}_{\scriptscriptstyle(#1)}} 
\begin{document}

\title{Quantum groups and interacting quantum fields}
\author{Christian Brouder}
\address{Laboratoire de Min\'eralogie-Cristallographie, CNRS UMR7590,
 Universit\'es Paris 6 et 7, IPGP, 4 place Jussieu,
  75252 Paris Cedex 05, France}
\begin{abstract}
If $C$ is a cocommutative coalgebra, a bialgebra structure can be given to
the symmetric algebra $S(C)$. The symmetric product is twisted by a Laplace
pairing and the twisted product of any number of elements of $S(C)$ is 
calculated explicitly. This is used to recover important identities in the 
quantum field theory of interacting scalar bosons.
\end{abstract}

\section{Introduction}
Quantum groups appear to be a powerful tool for quantum
field calculations. First, Fauser pointed out a connection 
between Wick's theorem and the concept of Laplace pairing 
introduced by Rota and his school \cite{Fauser}.
In reference \cite{BrouderOeckl}, we showed that the Laplace 
pairing is a quantum group concept and we solved quantum field 
problems with quantum group tools. The quantum group
approach to free scalar fields was presented in detail
in \cite{QFA}. Here we show that quantum groups 
can also deal with interacting fields.
In the first part, a bialgebra $S(C)$ is built from a cocommutative
coalgebra $C$, the symmetric product of $S(C)$ is 
twisted by a Laplace pairing and general identities are derived.
In the second part, these identities are translated into
the language of quantum field theory.

\section{The abstract setting}
If $C$ is a cocommutative coalgebra with coproduct
$\Delta'$ and counit $\counit'$, the symmetric
algebra $S(C)=\bigoplus_{n=0}^\infty S^n(C)$ 
can be equipped with the structure
of a bialgebra. 
The product of the bialgebra $S(C)$ is the symmetric product 
(denoted by $\cdot$) and
its coproduct $\Delta$ is defined on $S^1(C)=C$
by $\Delta a = \Delta' a$ and extended to
$S(C)$ by $\Delta 1=1\otimes 1$ and $\Delta (u\cdot v) =
\sum u\i1\cdot v\i1 \otimes u\i2\cdot v\i2$.
The elements of $S^n(C)$ are said to be of degree $n$.
The counit $\counit$ of $S(C)$ is defined to be
equal to $\counit'$ on $S^1(C)=C$ and extended to
$S(C)$ by $\counit(1)=1$ and $\counit(u\cdot v)=\counit(u)\counit(v)$.
It can be checked recursively that $\Delta$ is coassociative
and cocommutative
and that $\sum \counit(u\i1)u\i2=\sum u\i1\counit(u\i2)=u$.
Thus, $S(C)$ is a commutative and cocommutative bialgebra which 
is graded as an algebra.

A Laplace pairing is a bilinear map $(|)$ from $S(C)\times S(C)$
to the complex numbers such that $(1|u)=(u|1)=\counit(u)$,
$(u\cdot v|w)=\sum (u|w\i1)(v|w\i2)$ and 
$(u|v\cdot w)=\sum (u\i1|v)(u\i2|w)$
for any $u$, $v$ and $w$ in $S(C)$. 
The powers $\Delta^k$ of the coproduct 
are defined by $\Delta^0 a=a$,
$\Delta^1 a=\Delta a$ and
$\Delta^{k+1} a = (\Id\otimes\dots\otimes \Id\otimes\Delta)\Delta^k a$.
Their action is denoted by
$\Delta^k a =\sum a\i1\otimes\dots\otimes a\i{k+1}$.

From the definition of the Laplace pairing
and of the powers of the coproduct
a straighforward recursive proof yields, for
$u^i$ and $v^j$ in $S(C)$
\begin{eqnarray}
(u^1\cdot\dots\cdot u^k|v^1\cdot\dots\cdot v^l)
&=&
\sum \prod_{i=1}^k\prod_{j=1}^l (u\i{j}^i|v\i{i}^j).
\label{u1uk}
\end{eqnarray}
For example
$(u\cdot v\cdot w|s\cdot t)=\sum (u\i1|s\i1)(u\i2|t\i1)(v\i1|s\i2)(v\i2|t\i2)
        (w\i1|s\i3)(w\i2|t\i3)$.

A Laplace pairing is
entirely determined by its value on $C$. In other words,
once we know $(a|b)$ for all $a$ and $b$ in $C$,
equation (\ref{u1uk}) enables us to calculate the Laplace
pairing on $S(C)$.

The Laplace pairing induces a twisted product $\circ$ on $S(C)$
by $u\circ v=\sum (u\i1|v\i1) u\i2\cdot v\i2$. By applying the
counit to both sides of this equality we obtain
the useful relation \cite{QFA}
\begin{eqnarray}
\counit(u\circ v) &=& (u|v)
\label{counitucircv}
\end{eqnarray}
If we follow the proofs given in \cite{QFA} we can easily 
show that the twisted
product is associative, $1$ is the unit of $\circ$,
$(u\circ v|w)=(u|v\circ w)$
and $\Delta (u\circ v)=\sum u\i1\circ v\i1\otimes u\i2\cdot v\i2$.
If we use the last identity recursively, we obtain
for $u^1,\dots,u^k$ in $S(C)$
\begin{eqnarray}
\Delta (u^1\circ\dots\circ u^k) &=&
\sum u\i1^1\circ\dots\circ u\i1^k \otimes u\i2^1\cdot\dots\cdot u\i2^k.
\label{Deltaa1circak}
\end{eqnarray}

This leads us to the important relation
\begin{eqnarray}
u^1\circ\dots\circ u^k &=&
\sum \counit(u\i1^1\circ\dots\circ u\i1^k) u\i2^1\cdot\dots\cdot u\i2^k.
\label{a1circak}
\end{eqnarray}
To show (\ref{a1circak}) recursively, we observe that it is true
for $k=2$. We denote $U=u^1\circ\dots\circ u^k$
and we assume that the property is true up to the twisted product
of $k$ terms. Since, by definition,
$U\circ v=\sum (U\i1|v\i1) U\i2\cdot v\i2$,
equation (\ref{Deltaa1circak}) yields
$U\circ v=\sum (u\i1^1\circ\dots\circ u\i1^k|v\i1)
  u\i2^1\cdot\dots\cdot u\i2^k\cdot v\i2$
and the result follows for the twisted product of 
$k+1$ terms because of equation (\ref{counitucircv}).

Finally, we shall prove the second important identity
\begin{eqnarray}
\counit(u^1\circ\dots\circ u^k)
&=& \sum\prod_{i=1}^{k-1}\prod_{j=i+1}^{k}
  (u\i{j-1}^i|u\i{i}^{j})
= \sum\prod_{j>i} (u\i{j-1}^i|u\i{i}^{j}).
\label{epsa1circakr}
\end{eqnarray}
For $k=2$, equation (\ref{epsa1circakr}) is true because
of equation (\ref{counitucircv}). Assume that it is true
up to $k$ and denote $U=u^1\circ\dots\circ u^k$.
From equation (\ref{counitucircv}) and 
$U=\sum \counit(U\i1)U\i2$ we find
\begin{eqnarray*}
\counit(U\circ u^{k+1}) &=& \sum \counit(U\i1) (U\i2| u^{k+1})
=
\sum \counit(u\i1^1\circ\dots\circ u\i1^k) 
  (u\i2^1\cdot\dots\cdot u\i2^k| u^{k+1})
\\&=&
\sum \counit(u\i1^1\circ\dots\circ u\i1^k) 
   \prod_{n=1}^k (u\i2^n|u\i{n}^{k+1}),
\end{eqnarray*}
where we used equations (\ref{Deltaa1circak}) and 
(\ref{u1uk}).
Equation (\ref{epsa1circakr}) is true up to $k$ thus
\begin{eqnarray*}
\counit(u^1\circ\dots \circ u^{k+1}) &=& 
\sum\prod_{i=1}^{k-1}\prod_{j=i+1}^{k}
  (u\i{j-1}^i|u\i{i}^{j})
   \prod_{n=1}^k (u\i{k}^n|u\i{n}^{k+1})
\\&=&
\sum\prod_{i=1}^{k-1}\prod_{j=i+1}^{k+1}
  (u\i{j-1}^i|u\i{i}^{j})
   (u\i{k}^k|u\i{k}^{k+1})
=
\sum\prod_{i=1}^{k}\prod_{j=i+1}^{k+1}
  (u\i{j-1}^i|u\i{i}^{j})
\end{eqnarray*}
and the identity is proved for the twisted product of $k+1$ elements.

We considered the symmetric algebra $S(C)$, but the same
results are obtained for the tensor algebra $T(C)$.
A related construction was
made by Hivert in \cite{HivertPhD}.

\section{Quantum fields}
The previous construction is now applied to interacting quantum field
theory.
The scalar fields are defined by the usual formula
\cite{ReedSimonII}
\begin{eqnarray*}
\phi(x) &=& \int \frac{\dd\kbf}{(2\pi)^3\sqrt{2\omega_k}}
\Big(\ee^{-i p\cdot x} a(\kbf) + \ee^{i p\cdot x} a^\dagger(\kbf)\Big),
\end{eqnarray*}
where $\omega_k=\sqrt{m^2+|\kbf|^2}$, $p=(\omega_k,\kbf)$,
$a^\dagger(\kbf)$ and $a(\kbf)$ are the creation and annihilation
operators
acting on the symmetric Fock space of scalar particles.
Interacting fields are products of fields at the same point.
Thus, we define the powers of fields $\phi^n(x)$ as
the normal product of $n$ fields at $x$
(i.e.  $\phi^n(x)={:}\phi(x)\dots\phi(x){:}$).
This definition is meaningful for all $n>0$ and is extended
to $n=0$ by saying that $\phi^0(x)$ is the unit operator.
In the following we shall consider the divided powers of fields
defined by $\phi^{(n)}(x)=\phi^n(x)/n!$.

We consider the coalgebra $\calC$ generated by $ \phi^{(n)}(x)$,
where $x$ runs over spacetime and
$n$ goes from 0 to 3 for a $\phi^3$ theory and
from 0 to 4 for a $\phi^4$ theory.
We do not consider here the topology of this space.
The coproduct of $\calC$ is
$\Delta \phi^{(n)}(x) = \sum_{k=0}^n 
   \phi^{(k)}(x) \otimes  \phi^{(n-k)}(x)$
and its counit is $\counit( \phi^{(n)}(x))=\delta_{n,0}$.
Scalar fields are bosons, so we work with the symmetric
algebra $S(\calC)$. The product of $S(\calC)$ is
the normal product of operators, which is commutative
and denoted by ${:}uv{:}$. Notice that the
counit is equal to the expectation
value over the vacuum:
$\epsilon(u)=\langle 0|u|0\rangle$.

In $S(\calC)$, the Laplace pairing is entirely determined by the
value of $(\phi^{(n)}(x)|\phi^{(m)}(y))$,
which is itself
determined by the value of $(\phi(x)|\phi(y))=G(x,y)$
if we consider $\phi^{(n)}(x)$ as a product of fields.
More precisely
$(\phi^{(n)}(x)|\phi^{(m)}(y)) = \delta_{n,m} 
G(x,y)^{(n)}$, where the right hand side is a divided
power $G(x,y)^{(n)}=(1/n!) G(x,y)^n$. In general, $G(x,y)$ is
a distribution. In quantum field theory we use two special
cases: the Wightman function
$G_+(x,y)=\langle 0|\phi(x)\phi(y)|0\rangle$
and the Feynman propagator
$G_F(x,y)=\langle 0|T\big(\phi(x)\phi(y))|0\rangle$.
As shown in \cite{BrouderOeckl}, when
the Laplace pairing is defined with $G(x,y)=G_+(x,y)$, the
twisted product equals the operator product of fields.
When it is defined with $G(x,y)=G_F(x,y)$ the twisted product equals
the time-ordered product.

Notice that
$\Delta^{k-1} \phi^{(n)}(x)=\sum \phi^{(m_1)}(x)\otimes\dots\otimes
\phi^{(m_k)}(x)$, with a sum over all nonnegative integers
$m_i$ such that $\sum_{i=1}^k m_i=n$.
Thus, we can specialize equation (\ref{u1uk}) to our coalgebra $\calC$
\begin{eqnarray}
({:}\phi^{(n_1)}(x_1)\dots \phi^{(n_k)}(x_k){:}|
 {:}\phi^{(p_1)}(y_1)\dots \phi^{(p_l)}(y_l){:})
&=&
\sum_{M} \prod_{i=1}^k\prod_{j=1}^l G(x_i,y_j)^{(m_{ij})},
\label{Laplacen}
\end{eqnarray}
where the sum is over all $k\times l$ matrices $M$ of nonnegative 
integers $m_{ij}$ such that $\sum_{j=1}^l m_{ij}=n_i$ and
$\sum_{i=1}^k m_{ij}=p_j$. 
This formula was given in reference \cite{RotaStein}.

Equation (\ref{a1circak}) applied to $\calC$ yields
a classical result of quantum field theory
\begin{eqnarray}
\phi^{(n_1)}(x_1)\circ\dots \circ \phi^{(n_k)}(x_k) &=&
\sum_{i_1=0}^{n_1}\cdots\sum_{i_k=0}^{n_k}
\langle 0|\phi^{(i_1)}(x_1)\circ\dots \circ \phi^{(i_k)}(x_k)
|0\rangle 
\nonumber\\&&\hspace*{10mm}
{:}\phi^{(n_1-i_1)}(x_1)\dots \phi^{(n_k-i_k)}(x_k){:}.
\label{EG}
\end{eqnarray}
This equation was published 
by Epstein and Glaser for the operator product and the time-ordered 
product \cite{EpsteinGlaser}. It is now often used in the Epstein-Glaser
approach to renormalisation (see e.g. \cite{HollandsWald2}).
Equation (\ref{a1circak}) is clearly more compact and also
more general than equation (\ref{EG}): it is still valid
if the elements of $\calC$ (i.e. $\phi^{(n_i)}(x_i)$) are replaced by
elements of $S(\calC)$.

Finally, if we specialize equation (\ref{epsa1circakr}) to $\calC$
we obtain
\begin{eqnarray}
\langle 0|\phi^{(n_1)}(x_1)\circ\dots \circ \phi^{(n_k)}(x_k)
|0\rangle &=&
\sum_{M} \prod_{i=1}^{k-1}\prod_{j=i+1}^k G(x_i,x_j)^{(m_{ij})},
\label{omegab}
\end{eqnarray}
where the sum is over all symmetric $k\times k$ matrices $M$ of nonnegative 
integers $m_{ij}$ such that $\sum_{j=1}^k m_{ij}=n_j$ and
$m_{ii}=0$ for all $i$. 

When the twisted product is the operator product, this expression 
was given by Brunetti, Fredenhagen and K\"ohler \cite{BFK}. Notice 
that (\ref{omegab}) was proved here with a few lines of algebra,
whereas the quantum field proof is combinatorial.
As remarked by Rota, a great virtue
of Hopf algebras is to replace combinatorics by algebra.

When the twisted product is the time-ordered product, equation 
(\ref{omegab}) has a diagrammatic interpretation. 
The diagrammatic calculation of 
$\langle 0 |T\big(\phi^{(n_1)}(x_1)\dots \phi^{(n_k)}(x_k)
\big) |0\rangle$
would be: draw all diagrams that have $k$ vertices $x_1$ to $x_k$
and for which each vertex $x_i$ has $n_i$ edges.
There is a one to one correspondence
between these diagrams and the matrices $M$ satisfying the conditions
stated above: $m_{ij}$ is the number of edges linking vertices 
$x_i$ and $x_j$. The condition $m_{ii}=0$ means that there is no 
tadpoles.
The graphs are not directed (i.e. the edges do not carry arrows)
because the Feynman propagator $G_F(x,y)$ is symmetric
(i.e. $G_F(y,x)=G_F(x,y)$).

\section{Perspective}
This paper shows that non trivial results of quantum field
theory can be derived easily from a general quantum group 
construction. A word of caution must be added concerning
equation (\ref{omegab}). 
When the twisted product is the operator product,
equation (\ref{omegab}) is valid. It defines a state
on $T(\calC)$ by 
$\omega(a_1\otimes\dots\otimes a_k)=\epsilon(a_1\circ\dots\circ a_k)$
and the Laplace pairing is a positive semidefinite form on
$T(C)\times T(C)$.
However, when the twisted product is the time-ordered product, 
equation (\ref{omegab}) is ill-defined because the
powers $G_F(x,y)^n$ are singular products of distributions
and renormalisation is necessary.
The first step of a ``quantum group'' renormalisation of scalar field
theories was done in \cite{BrouderOeckl}.
It uses the fact that equation (\ref{EG}) is still
valid in renormalised quantum field theory, so that
the Laplace pairing must be replaced by a Sweedler's
2-cocycle \cite{Majid} in the definition of a renormalised
time-ordered product.
To go further, we can implement renormalisation abstractly
by starting from a bialgebra $B$ and putting a bialgebra
structure on the ``squared'' tensor algebra $T(T(B)^+)$
\cite{HivertPhD}.
This construction is inspired by Pinter's approach
to renormalization \cite{PinterHopf} and is related
to the Fa{\`a} di Bruno bialgebra of composition
of series \cite{Majid}. These results will be presented 
in a forthcoming publication.

\section{Acknowledgements}
I am very grateful to Robert Oeckl for his comments on
this paper and to Alessandra Frabetti for her enthusiastic support. 
This is IPGP contribution \#0000.


\end{document}